\RequirePackage{lineno} 
\documentclass[article]{revtex4}

\usepackage{graphicx}% Include figure files
\usepackage{dcolumn}% Align table columns on decimal point
\usepackage{bm}% bold math
\usepackage{color}
\usepackage{eso-pic}
\usepackage{amsmath}
 \usepackage[retainorgcmds]{IEEEtrantools}
\usepackage{xspace}
\newcommand{\LTO}{$\mathrm{LaTiO}_3$\xspace}
\newcommand{\STO}{$\mathrm{SrTiO}_3$\xspace}
\newcommand{\LAO}{$\mathrm{LaAlO}_3$\xspace}
\newcommand{\TiO}{$\mathrm{TiO}_2$\xspace}
\newcommand{\SN}{$\mathrm{Si}_3\mathrm{N}_4$\xspace}
\newcommand{\YBCO}{$\mathrm{YBa}_2\mathrm{Cu}_3\mathrm{O}_7$\xspace}
\newcommand{\LXO}{$\mathrm{LaXO}_3\mathrm{(X=Al,Ti)}$\xspace}
\begin{document}

\title{Field-effect control of superconductivity and Rashba spin-orbit coupling in top-gated \LAO/\STO devices}

\author{S. Hurand$^1$, A. Jouan$^1$, C. Feuillet-Palma$^1$, G. Singh$^1$, J. Biscaras$^1$, E. Lesne$^2$, N. Reyen$^2$, A. Barth\'el\'emy$^2$, M. Bibes$^2$, C. Ulysse$^3$, X. Lafosse$^3$, M. Pannetier-Lecoeur$^4$, S. Caprara$^5$, M. Grilli$^5$, J. Lesueur$^1$, N. Bergeal$^1$}

\affiliation{$^1$Laboratoire de Physique et d'Etude des Mat\'eriaux -CNRS-ESPCI ParisTech-UPMC, PSL Research University, 10 Rue Vauquelin - 75005 Paris, France.}
\affiliation{$^2$Unit\'e Mixte de Physique CNRS-Thales, 1 Av. A. Fresnel, 91767 Palaiseau, France }
\affiliation{$^3$Laboratoire de Photonique et de Nanostructures LPN-CNRS, Route de Nozay, 91460 Marcoussis, France}
\affiliation{$^4$DSM/IRAMIS/SPEC - CNRS UMR 3680, CEA Saclay, F-91191 Gif sur Yvette Cedex, France}
\affiliation{$^5$Dipartimento di Fisica Universit\`{a} di Roma``La Sapienza'', piazzale Aldo Moro 5, I-00185 Roma, Italy}

\date{\today}

\begin{abstract}

The recent development in the fabrication of artificial oxide heterostructures opens new avenues in the field of quantum materials by enabling the manipulation of the charge, spin and orbital degrees of freedom. In this context, the discovery of two-dimensional electron gases (2-DEGs) at  \LAO/\STO interfaces, which exhibit both superconductivity and strong Rashba spin-orbit coupling (SOC), represents a major breakthrough. Here, we report on the realisation of a field-effect \LAO/\STO  device, whose physical properties, including superconductivity and SOC, can be tuned over a wide range by a top-gate voltage. We derive a phase diagram, which emphasises a field-effect-induced superconductor-to-insulator quantum phase transition.  Magneto-transport measurements indicate that the Rashba coupling constant increases linearly with electrostatic doping. Our results pave the way for the realisation of mesoscopic devices, where these two properties can be manipulated on a local scale by means of top-gates.
\end{abstract}

\maketitle

The interplay between superconductivity and spin-orbit coupling (SOC) is at the centre of intensive research efforts as it can generate a variety of unique phenomena such as the occurrence of triplet superconductivity, for instance \cite{gorkov}. Recently, hybrid nanostructures involving a superconductor in proximity to a semiconducting nanowire with a strong SOC have been proposed as an ideal system to observe a topological superconducting phase, which accommodates pairs of Majorana fermions \cite{lutchyn,oreg}. Following this idea, the first signatures of Majorana Fermions were obtained in devices made with indium antimonide in contact with niobium titanium nitride \cite{mourik}. However, the realisation of such devices remains a challenge because (i) the intrinsic value of the SOC in semiconductors is weak and cannot be tuned (ii) it is difficult to control the spin state at the interface between very different materials. For this reason, the discovery of a two-dimensional electron gas (2-DEG) at the interface between two insulating oxides such as \LAO/\STO or \LTO/\STO raised a considerable interest \cite{Ohtomo:2004p442}.  Indeed, this 2-DEG displays both superconductivity \cite{Reyren:2007p214,Biscaras:2010p7764} and strong Rashba SOC \cite{caviglia2, benshalom}, a combination of properties which is rarely observed in the same material.  \\

  The 2-DEG whose typical extension in the  \STO substrate is of order $\sim$10 nm \cite{Copie:2009p5635,biscaras2} is confined in an interfacial quantum well buried under an   few unit cells thick insulating \LAO  layer. By adjusting the Fermi level with a gate voltage, the conductivity of the 2-DEG can be modulated from insulating to superconducting \cite{Caviglia:2008p116,biscaras2}. In addition, the Rashba SOC, which is dominated by the local electric field at the interface, can be increased by filling the quantum well \cite{caviglia2}. The combination of these two effects enables the realisation of  nanostructures, where the very same material can be turned into different states by applying a local electric field-effect.  Thus far,  controlling the superconductivity and SOC have been demonstrated almost exclusively with  gates deposited at the back of thick \STO substrates. Because of the very high value of the \STO dielectric constant at low temperatures ($\epsilon\approx$ 20000) \cite{NEVILLE:1972p3397}, the electric field-effect can  significantly modulate the carrier density with gate voltages on the order of 100 V \cite{Bell:2009p6086,Caviglia:2008p116,biscaras2}. However, in such geometry, it is not possible to control the properties of the 2-DEG on a scale much smaller than the typical thickness of the substrate (500 $\mu$m), making it impossible to realise devices with dimensions comparable to lengths that are characteristic of quantum orders (such as the superconducting coherence length and the spin diffusion length). To overcome this problem, field-effect control of the superconductivity and Rashba SOC needs to be achieved by means of local top-gates.  Forg et al. fabricated  field-effect transistors in a \LAO/\STO heterostructures using the insulating \LAO layer as the gate dielectric and the \YBCO layer as the top-gate electrode \cite{forg}. Hosoda et al.  achieved top-gate control of the normal state properties using a metallic gate directly deposited on the \LAO layer \cite{hosoda}.  More recently, a first attempt to modulate the superconductivity with a top-gate gave promising results \cite{eerkes}, despite the leaky insulating \LAO layer. In this article, the realisation of a top-gated field-effect device is reported. The properties of the 2-DEG could be tuned over a wide range, from a superconducting to an insulating state. In addition, the control of the Rashba SOC by means of a top-gate is also demonstrated. \\

 A ten-$\mu m$-wide superconducting Hall bar was first fabricated with an amorphous \LAO template method and then covered by a \SN dielectric layer and a metallic top gate  (see Fig. 1a and 1b) \cite{stornaiuolo}.  More information on the fabrication processes is given in the Methods section. The sample was anchored to the mixing chamber of a dilution refrigerator with a base temperature of 16 mK. Figure 1c shows the superconducting transition of the device at the critical temperature  T$_c^{\mathrm onset}\approx$ 250 mK, which is similar to an unprocessed \LAO/\STO heterostructure. The current-voltage (I-V) characteristics of the device  abruptly switches from the superconducting state ($R=0$) to the resistive state ($R\neq 0$) at the critical current $I_c$= 460 nA which corresponds to a critical current density of approximately 500 $\mu$A/cm. \\

\textbf{Electrostatic control of the carrier density}\\

	After the sample was cooled, the top-gate voltage $V_\mathrm{TG}$ was first increased to  +110 V, beyond the saturation threshold of the resistance.  During this operation,  electrons are added in the quantum well, increasing the Fermi energy to its maximum value (i.e., the top of the well) \cite{biscaras3}. In comparison with back-gate experiments where the relationship between the carrier density ($n$) and the back-gate voltage $V_\mathrm{BG}$ is not trivial owing  to the electric-field-dependent dielectric constant of \STO \cite{NEVILLE:1972p3397}, here, the carrier density is expected to increase linearly with  $V_\mathrm{TG}$.  Figure 2 shows the sheet carrier density $n=\frac{-IB}{eV_H}$, extracted from the Hall effect measurements performed up to B=4 T as a function of the top-gate voltage $V_\mathrm{TG}$, for two different back-gate voltages ($V_\mathrm{BG}$=0 V and $V_\mathrm{BG}$=-15 V). For $V_\mathrm{BG}$=0 V,  the linear increase in $n$ is observed with $V_\mathrm{TG}$ only for negative $V_\mathrm{TG}$. The non-physical decrease in $n$ with $V_\mathrm{TG}$ for positive gate voltages is caused by the incorrect determination of the carrier density at low magnetic fields. It was shown that at the \LAO/\STO interface, the Hall voltage is no longer linear with the magnetic field for strong filling of the quantum well because of multi-band transport  \cite{Kim:2010p9791,Ohtsuka:2010p9619,biscaras2}.  To reach a doping regime where the one-band approximation is valid, a negative back-gate $V_\mathrm{BG}$=-15 V was applied producing a depletion  of the highest energy sub-bands that accommodate the highly-mobile carriers, responsible of the decrease of the Hall number at positive $V_\mathrm{TG}$. Figure 2 shows that in this case, the linear dependence of $n=\frac{-IB}{eV_H}$ with $V_\mathrm{TG}$ can be recovered. The linear fit of slope $\frac{dn}{dV_\mathrm{TG}}$=5.0~$\times$~$10^{10}$ $e^{-}\cdot\mbox{cm}^{-2}\cdot$V$^{-1}$ is obtained from numerical simulations of the electric field-effect by a finite elements method assuming a dielectric constant  $\epsilon$=5 for the \SN layer (see the inset in Fig. 2). Finally,  the following relationship between the carrier density and top-gate voltage is deduced: $n$=5.0~$\times$~$10^{10}$ $V_\mathrm{TG}$+1.69~$\times$~$10^{13}$ $e^{-}\cdot\mbox{cm}^{-2}$.\\
	
\textbf{Superconductivity and phase diagram}\\
	
	In the following, the back gate voltage $V_\mathrm{BG}$ was always set to 0 V unless otherwise stated. Figure 3a shows the sheet resistance of the device as a function of temperature measured for different top-gate voltages in the  range [-110 V,+110 V], where the leakage gate current is negligible ($<$0.1 nA).  The variation in $V_\mathrm{TG}$ induces a modulation in the normal state resistance by two orders of magnitude.  Figure 3b summarises the variations of the normalised resistance $R/R$($T$=350 mK)  as a function of temperature ($T$) and top-gate voltage $V_\mathrm{TG}$ on a phase diagram. The corresponding $n$ is also indicated on the top axis. The device displays a gate-dependent superconducting transition, whose critical temperature $T_c$ describes a partial dome  as a function of $V_\mathrm{TG}$, similar to that observed with a back-gate \cite{Bell:2009p6086,Caviglia:2008p116,biscaras2}.   The maximum $T_c$, corresponding to optimal doping, is around 250 mK. In the underdoped region, a decrease in the gate voltage causes  $T_c$ to continuously decrease from its maximum value to zero.  A superconductor-to-insulator quantum phase transition takes place around $V_\mathrm{TG}$=-90 V. The critical sheet resistance at the transition is $R_s\simeq$ 8 k$\Omega$, which is close to the quantum of resistance of bosons with 2e charges, $R_Q=\frac{h}{4e^2}\simeq$ 6.5 k$\Omega$. For large negative voltages, corresponding to low electron densities, the sheet resistance increases strongly when approaching the insulating state. In the overdoped region, the addition of electrons into the quantum well with the top-gate produces a small decrease in $T_c$ whose origin is currently under debate. Such behaviour has also been observed in doped bulk \STO \cite{takada} and could be reinforced by the two-dimensionality of the interface \cite{klimin}. The current-voltage characteristics of the device for different top-gate voltages are shown in Supplementary Material.   \\
	
\textbf{Rashba Spin-orbit coupling}\\

In \LAO/\STO heterostructures, the accumulation of electrons in the interfacial quantum well generates a strong local electric field $E_z$ perpendicular to the motion of the electrons, which translates into a magnetic field in their rest frame. The coupling of the electrons spins to this field gives rise to the so-called Rashba SOC \cite{rashba}, characterised by the  spin-splitting energy  $\Delta_\mathrm{SO}=2\alpha k_\mathrm{F}$ at  the Fermi surface,  where $k_F$ is the Fermi wave vector and  $\alpha$ is the Rashba coupling constant. This latter is directly proportional to the interfacial electric field $E_z$ and therefore depends on the filling of the quantum well. In electronic transport measurements, the presence of a spin-orbit coupling results in an additional spin relaxation mechanism characterised by the relaxation time $\tau_\mathrm{SO}$. The weak localization corrections to the conductance of a two-dimensionnal system at low temperatures is modified by $\tau_\mathrm{SO}$ \cite{maekawa,hikami}. The strength of the SOC can therefore be determined  by properly analysing the magnetoconductance $\Delta\sigma(B)=\sigma(B)-\sigma(0)$.\\
\indent  $\Delta\sigma(B)$ was measured in the normal state at different temperatures and  top-gate voltages. For negative $V_\mathrm{TG}$ a positive magnetoconductance was observed beyond 1T. This is characteristic of a weak localization regime with small SOC (Fig. 4). As $V_\mathrm{TG}$ is increased, an inversion of the sign of the magnetoconductance is observed and at large positive gate voltage the magnetoconductance remains always negative.The experimental data in Figure 4 were fitted with the Maekawa-Fukuyama formula in a diffusive regime that describes the change in the  conductivity with magnetic field with negligible Zeeman splitting \cite{maekawa}, 
 
\begin{IEEEeqnarray}{rCl} 
\Delta\sigma(B)/G_0&=&-\Psi\Big(\frac{1}{2}+\frac{B_\mathrm{tr}}{B}\Big)+\frac{3}{2}\Psi\Big(\frac{1}{2}+\frac{B_\Phi+B_{SO}}{B}\Big)\nonumber\\ 
&&-\frac{1}{2}\Psi\Big(\frac{1}{2}+\frac{B_\Phi}{B}\Big)-\Big[\ln\Big(\frac{B_\Phi+B_{SO}}{B_\mathrm{tr}}\Big)\nonumber\\
&&+\frac{1}{2}\ln\Big(\frac{B_\Phi+B_{SO}}{B_{_\Phi}}\Big)\Big]-A_K\frac{\sigma_0}{G_0}B^2
\label{MF}
  \end{IEEEeqnarray}
  
\noindent where $\Psi$ is the digamma function, $G_0=\frac{e^2}{\pi h}$ is the quantum of conductance, and the parameters  $B_\mathrm{tr}$, $B_\Phi$, $B_\mathrm{SO}$ are the effective fields related to the elastic, inelastic and spin-orbit relaxation times respectively. $B_\Phi$ and $B_\mathrm{SO}$, which are measured here by a transport experiment, are related to the relaxation times $\tau_\Phi$ and $\tau_\mathrm{SO}$ by the expressions $B_\Phi=\hbar/(4eD\tau_\Phi)$ and $B_\mathrm{SO}=\hbar/(4eD\tau_\mathrm{SO})$ respectively, where $D$ is the diffusion constant \cite{maekawa,hikami}. Finally, to account for the orbital magnetoconductance, we have added  in Eq. (1) a $B^2$ term with a Kohler coefficient  $A_K$ which increases quadratically with the mobility \cite{sommerfeld,macdonald}. Good agreement is obtained between the experimental data and the theory over the whole electrostatic doping range.  The evolution of the fitting parameters as a function of the top-gate voltage and equivalent carrier density is shown in Fig. 4b.  $B_\mathrm{\phi}$  varies only weakly over the whole range of gate voltage, indicating that the number of inelastic collisions does not depend on the carrier density. \\
 \indent In the framework of the weak localisation theory the temperature dependence of the inelastic scattering time is given by $\tau_\Phi\propto T^{-p}$ and therefore $B_\Phi\propto T^{p}$, where $p$ depends on the inelastic mechanism. The same fitting procedure was performed at different temperatures, giving a linear relationship between $B_\Phi$ and $T$ (Fig. 4b inset and Supplementary Material). This is consistent with $p$=1, which indicates that the inelastic scattering is dominated by electron-electron interactions \cite{lee, Biscaras:2010p7764}. \\
 
 \textbf{Spin-splitting energy.}\\
  
The spin-orbit term ($B_\mathrm{SO}$)  increases with top-gate voltage and, correspondingly, with the carrier density. The analysis of this dependence can shed light on the origin of the SOC at the \LAO/\STO interface.  If the spin relaxation is dominated by the D'Yakonov-Perel mechanism, based on a Rashba band splitting, $\tau_\mathrm{SO}=\frac{\hbar^4}{4\alpha^2Dm^2}$ in the limit $2\alpha k_F\tau_e/h\ll$1 ($\tau_e$ is the elastic relaxation time), which is satisfied here  \cite{dyakonov,hikami}. We then obtain the relationship between the coupling constant and the spin-orbit effective field  $\alpha=(e\hbar^3B_\mathrm{SO})^{1/2}/m$. Integrating the Maxwell-Gauss equation in the direction perpendicular to the interface gives $E_z=\frac{e}{\epsilon}(n+n_t)$ where $\epsilon$ is the dielectric constant of \SN at the interface and $n_t$ is the carrier density of non-mobile charges trapped in the \STO substrate. It is therefore expected that the Rashba coupling constant will vary with carrier density with the form $\alpha=an+b$, which is well satisfied experimentally for a wide range of electrostatic doping (Fig. 5). This indicates that the D'Yakonov-Perel mechanism in the presence of Rashba SOC is dominant in these 2-DEGs.  Assuming a Fermi energy of 100 meV and an effective mass $m=0.7m_0$ for the $d_{xy}$ light subbands mainly occupied, we can estimate the spin-splitting energy $\Delta_\mathrm{SO}=2k_F\alpha$. The order of magnitude of a few meV, which is much larger than in most semiconductors, is in agreement with previous studies \cite{caviglia2,benshalom}. Neglecting the small changes in $k_F$ with doping, we can plot the variation of $\Delta_\mathrm{SO}$ with $V_\mathrm{TG}$, and correspondingly, $n$ (Figure 5).  $\Delta_\mathrm{SO}$ is independent of the temperature below 10 K  as the shape of the quantum well and $E_z$ do not change in this temperature range (see the inset in Fig. 5 and Supplementary Material). The Kohler term (parameter $A_K$) is proportional to the square of the mobility. Thus, it dominates the magnetoconductance for gate voltages beyond +80 V, making it difficult to precisely evaluate the SOC at strong electrostatic doping. \\

In summary,  \LAO/\STO-based field-effect devices were fabricated using the amorphous \LAO template method. The  superconductivity can be electrostatically modulated over a wide range by a top-gate voltage, without any leakage. A superconductor-to-insulator quantum phase transition is induced when the quantum well is strongly depleted. By analysing the magnetotransport measurements, the presence of strong spin-orbit coupling that could be controlled with the top-gate voltage was demonstrated. The spin-spliting energy on the order of a few meV was found to increase linearly with the interfacial electric field in agreement with the Rashba mechanism. These results represent an important step toward the realisation of new mesoscopic devices, where the interaction between superconductivity and the Rasba SOC could give rise to non-conventional electronic states. \\

\textbf{Methods}\\

\indent \textbf{Device fabrication.} Starting with a \TiO-terminated $(0\,0\,1)$-oriented \STO commercial substrate  (Crystec), the template of a Hall bar  with contact pads was defined by evaporating an amorphous \LAO layer through a resist patterned by optical lithography. After a lift-off process, a thin layer of  crystalline \LAO (8 u.c) was grown on the amorphous template by Pulse Laser Deposition,  such that only the areas directly in contact with the substrate (Hall bar and contact pads) were crystalline. A KrF excimer (248\,nm) laser was used to ablate the single-crystalline \LAO target at 1\,Hz, with a fluence between 0.6 and 1.2\,J/cm$^2$ under an O$_2$ pressure of $2\times 10^{-4}$\,mbar \cite{edouard}. The substrate was typically kept at 650$^\circ$C during the growth of the film, monitored in real-time by reflection high-energy electron diffraction RHEED. As the growth occurs layer-by-layer,  the thickness can be controlled at the unit cell level. After the growth of the film, the sample was cooled down to 500$^\circ$C under a O$_2$ pressure of $10^{-1}$\, mbar, which was increased up to 400\,mbar.  To reduce the presence of oxygen vacancies (in both the substrate and the film), the sample was kept under these conditions for 30 minutes before it was cooled to room temperature. The 2-DEG forms at the interface between the crystalline \LAO layer and the \STO substrate. Such method has already been used to fabricate ungated 500 nm wide channels without noticeable alteration of the 2DEG properties \cite{stornaiuolo}. Once the channel is defined, a 500 nm thick \SN dielectric layer was deposited on the Hall bar by a lift-off process. After this step, a gold top-gate layer was deposited and lifted-off forming and appropriate geometry to cover the Hall bar. A metallic back gate was added at the end of the process.\\

\newpage

\noindent\textbf{\large{Acknowledgments}}\\
\normalsize

This work was supported by the french ANR, the DGA the CNRS PICS program and the R\'egion Ile-de-France through CNano IdF and Sesame programs.\\

Correspondence and requests for materials
should be addressed to N.B. (nicolas.bergeal@espci.fr)

\newpage

          \begin{figure}[h]
\includegraphics[width=12cm]{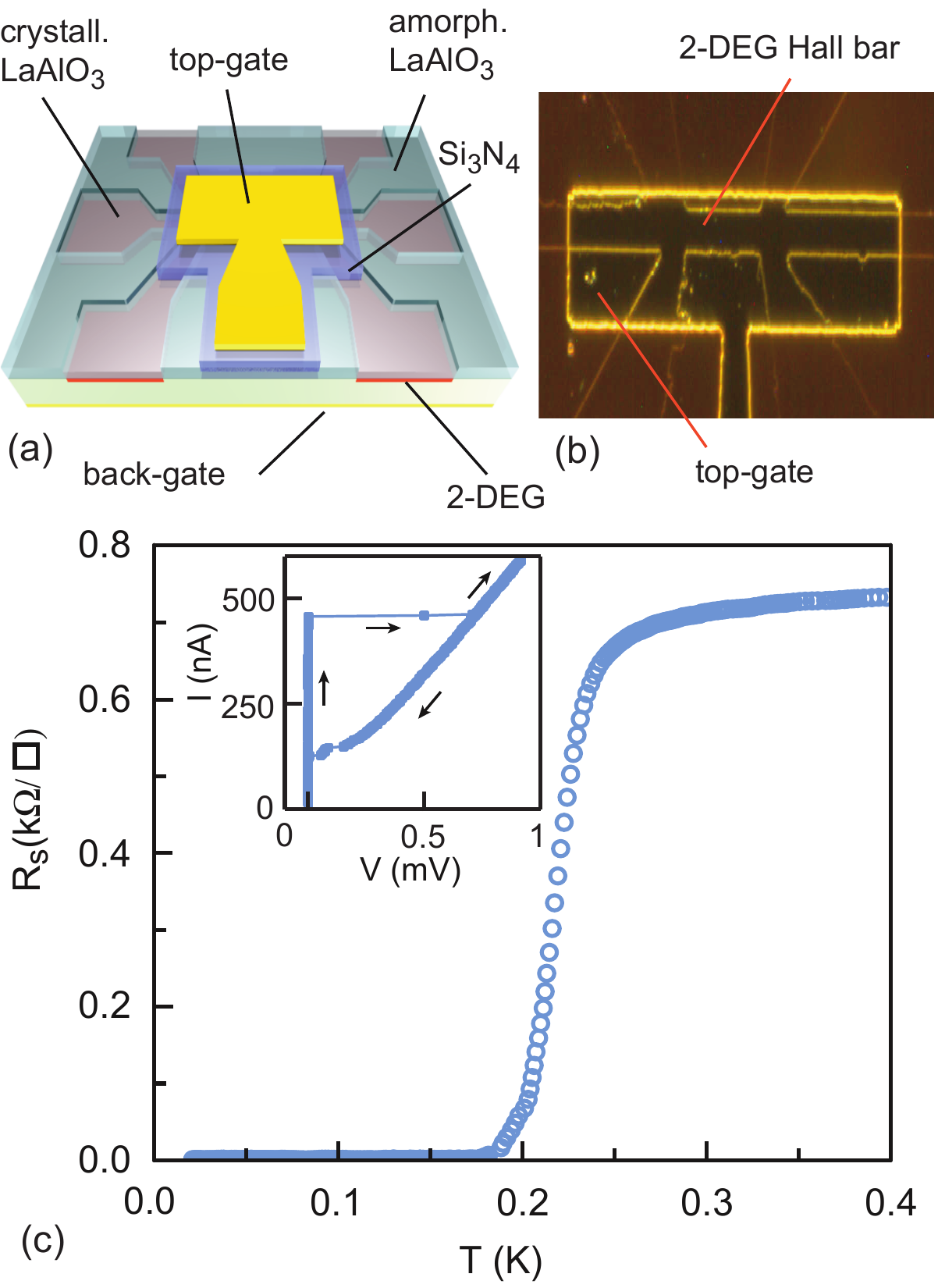}
\caption{ Device structure and superconducting transition. a) Schematic of the \LAO/\STO device with a 500 nm thick \SN dielectric layer. b) Dark-field optical picture of the device showing the Hall bar covered by a  top-gate. c) Sheet resistance as a function of temperature showing a superconducting transition at a critical transition temperature $T_c\approx$ 250 mK. Inset) Current-voltage characteristics of the device indicating the critical current $I_c$=460 nA. The arrows indicate the direction of the current sweep.\\}
\end{figure}

\newpage
          \begin{figure}[h]
\includegraphics[width=12cm]{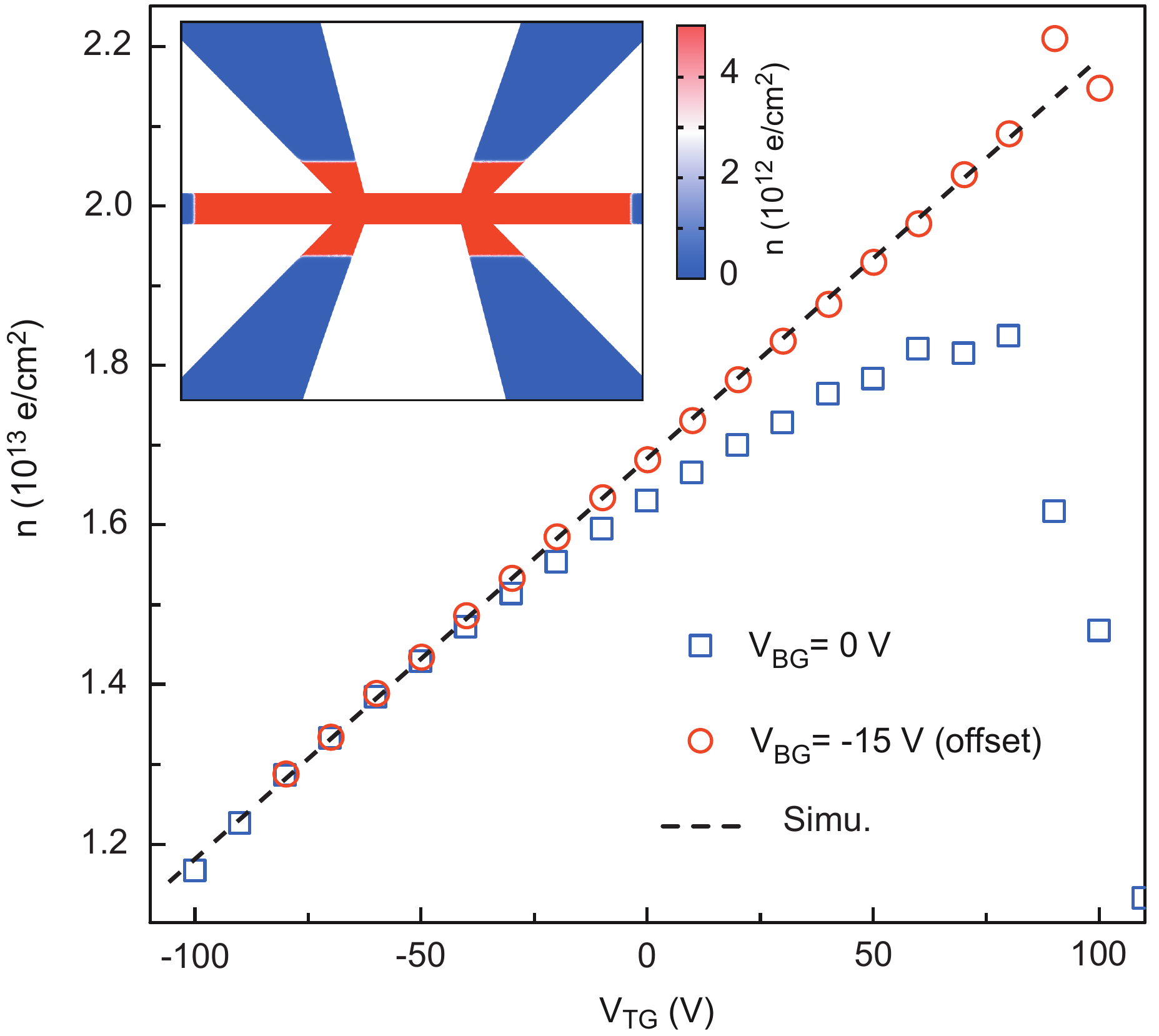}
\caption{ Hall effect and carrier density. Carrier density ($n$) extracted from the slope of the Hall voltage ($V_H$) at 4 T as a function of $V_\mathrm{TG}$ for two different  back-gate voltages ($V_\mathrm{BG}$). The curve at $V_\mathrm{BG}$= -15 V is offset to match the curve at  $V_\mathrm{BG}$= 0 V at negative top-gate voltages. The dashed line was obtained from numerical simulations on the carrier density, assuming a dielectric constant $\epsilon$=5 for the \SN layer. Inset: example of a numerical simulation of the charge carrier distribution in the device for $V_\mathrm{BG}$=0 V and $V_\mathrm{TG}$=10 V.}
\end{figure}

          \begin{figure}[h]
\includegraphics[width=12cm]{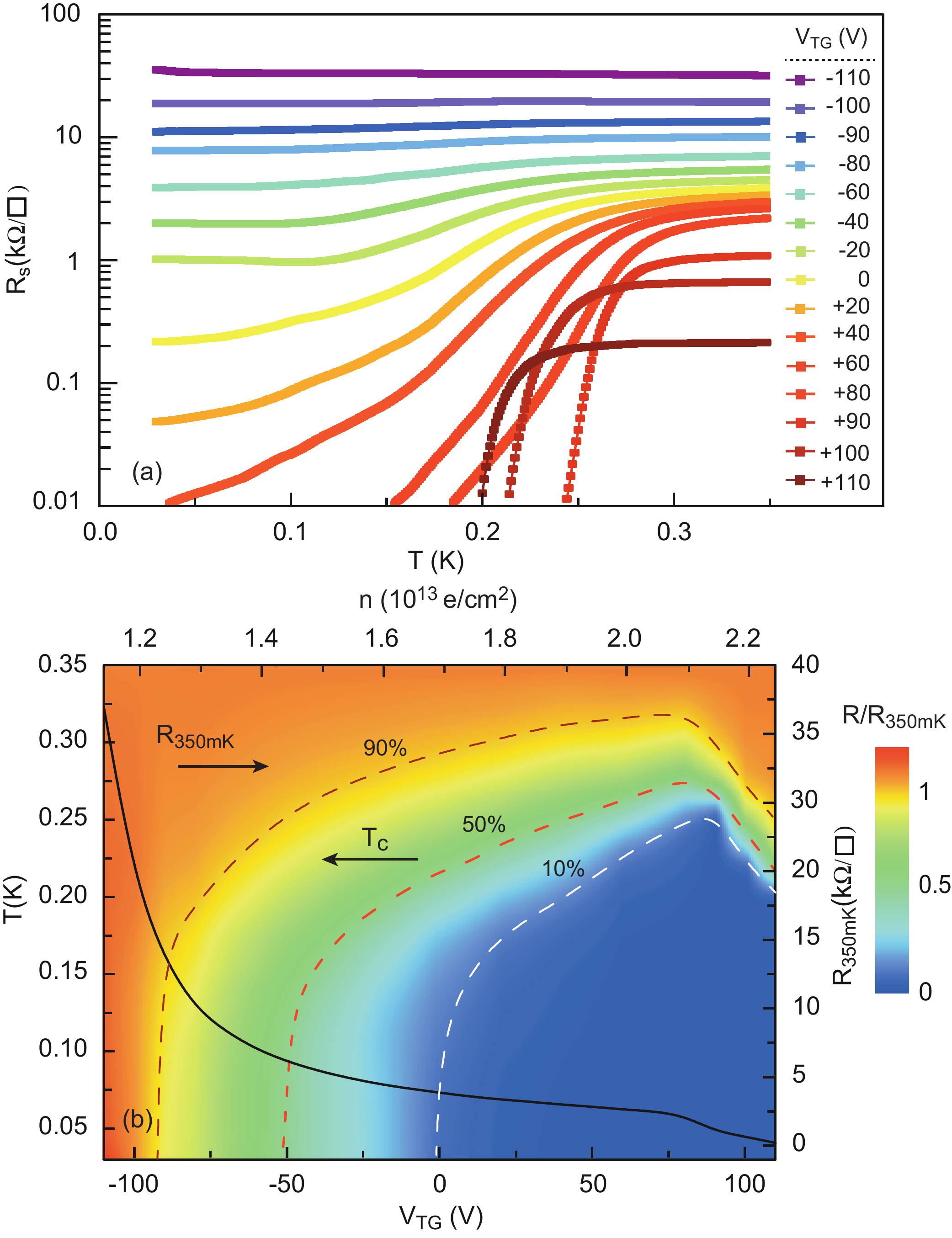}
\caption{ Field-effect control of the superconductivity. a) Sheet resistance of the device as a function of temperature for different $V_\mathrm{TG}$.  b) Sheet resistance normalised by its value at T=350mK plotted with a colour scale as a function of temperature (left axis) and top-gate voltage. The carrier densities corresponding to the top-gate voltages have been added in the top axis. The sheet resistance at T=350 mK is plotted as a function of top-gate voltage on the right axis. The critical temperature T$_c$  is plotted as function of the top-gate voltage on the left axis for the different criteria : drop of 10$\%$, 50$\%$ and 90 $\%$ of the normal resistance taken at T=350 mK.}
\end{figure}

          \begin{figure}[h]
\includegraphics[width=12cm]{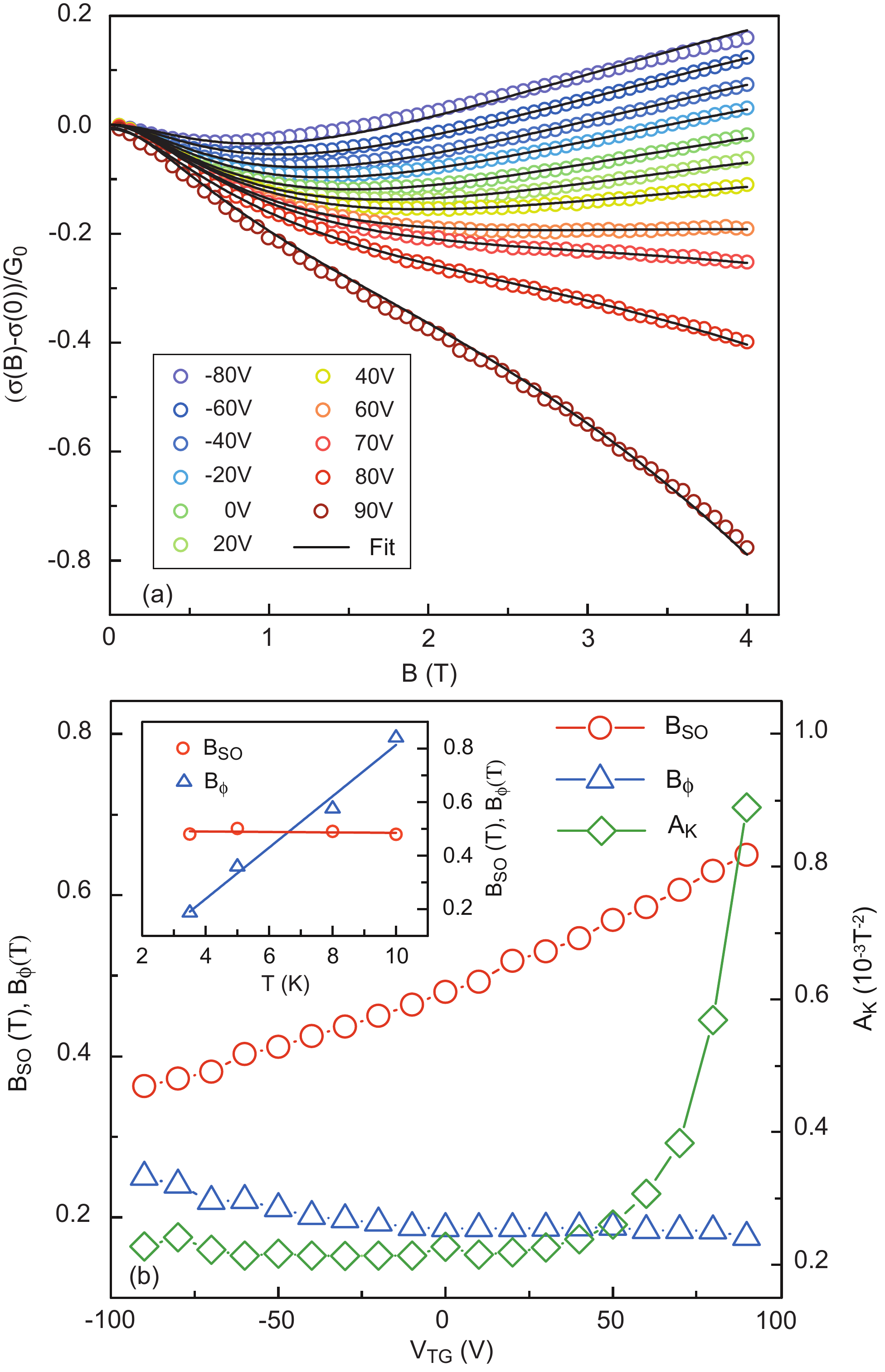}
\caption{Magnetotransport measurements. a) Magnetoconductance of the device at T=3.5 K for different ${V_\mathrm{TG}}$. The experimental data (open symbols) are fitted with the Maekawa-Fukuyama formula (\ref{MF}). b) Evolution of the fitting parameters $B_\mathrm{SO}$, $B_\mathrm{\phi}$ and $A_K$ as a function of the gate voltage. Inset) Variations in $B_\mathrm{SO}$ and $B_\mathrm{\phi}$ as a function of temperature for ${V_\mathrm{TG}}$=0.}
\end{figure}

          \begin{figure}[h]
\includegraphics[width=12cm]{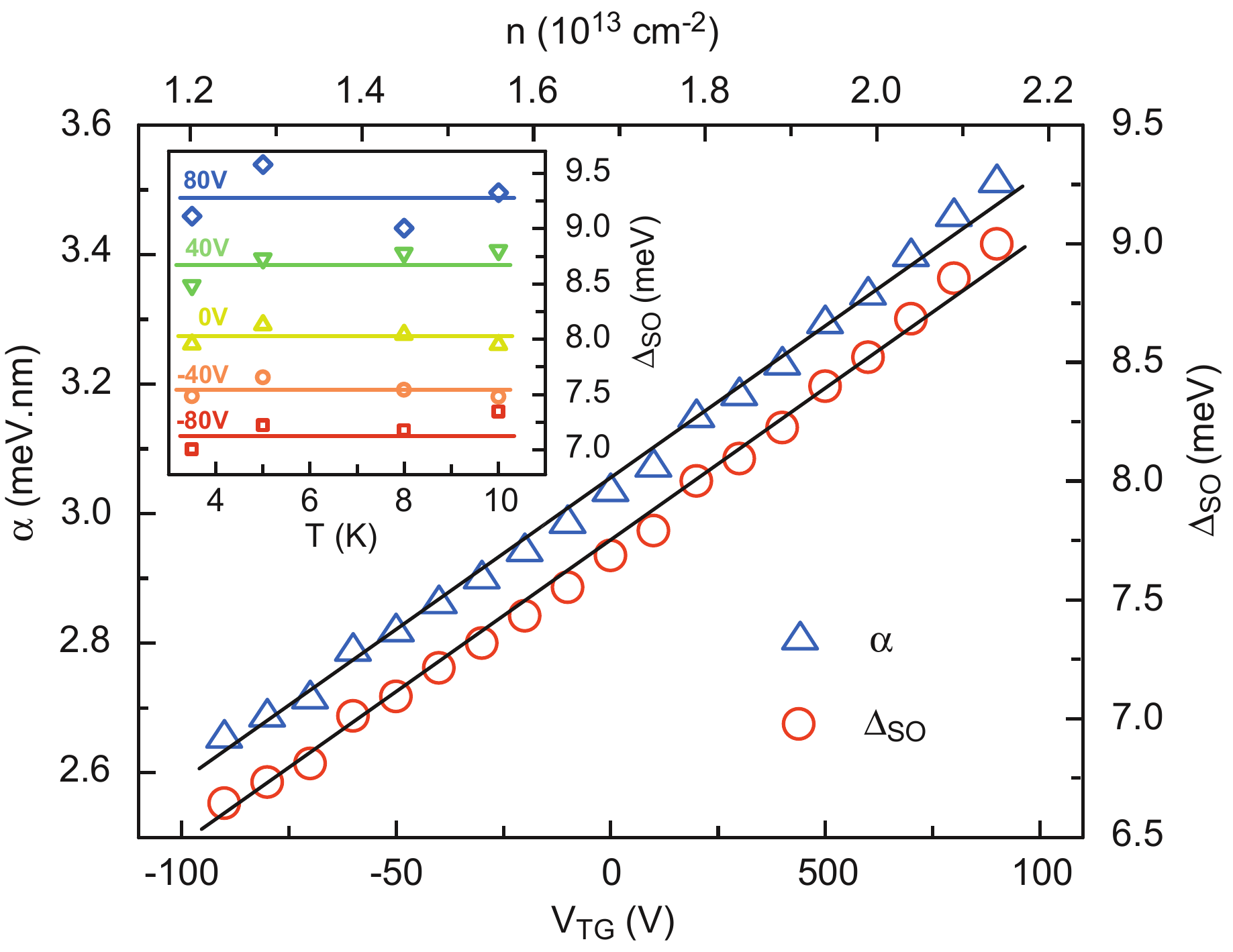}
\caption{Spin-orbit splitting. Spin-orbit constant and splitting as a function of $V_\mathrm{TG}$ (bottom axis) and corresponding carrier density (top axis). Inset :  Spin-orbit splitting as a function of temperature for selected ${V_\mathrm{TG}}$.}
\end{figure}

 \end{document}